\documentclass[a4paper,11pt]{article}
\usepackage{pos}
\usepackage{siunitx}
\usepackage{lineno}
\usepackage{url}

\title{The TOTEM nT2 detector: architecture, operation and performance}
\author*[a]{E. Bossini}

\affiliation[a]{Istituto Nazionale Fisica Nucleare, Sezione di Pisa,\\
 Largo B. Pontecorvo 3, Pisa, Italy}
 \onbehalf{for the TOTEM collaboration}

\emailAdd{edoardo.bossini@pi.infn.it}

\abstract{The TOTEM experiment at the LHC has produced a
large set of measurements on diffractive processes and pp cross sections.  
A new detector, called nT2, has been designed to measure the inelastic scattering rate during the LHC special run of 2023.
Due to the high radiation environment, 
the detector had to be installed in 10-20 minutes at most, then commissioned and operated after only few days.
The detector, based on plastic scintillators read out by matrices of SiPMs, was designed with such constraints in mind.
The front-end, DAQ and control electronics was developed with a fault tolerant architecture, moving as many functionalities as 
possible on a radiation tolerant SoC FPGA, hosting an integrated ARM controller.
Here we will describe the nT2 detector and its read-out and control electronics.
The detector was successfully operated during the special run: 
we will report the preliminary results on the detector performance.
}

\FullConference{42nd International Conference on High Energy Physics (ICHEP2024)\\
18-24 July 2024\\
Prague, Czech Republic\\}

\begin{document}
\maketitle

\section{Introduction}\label{sec:intro}
The TOTEM \cite{TOTEM_std} experiment is located at the Interaction Point (IP) 5 of the Large Hadron Collider (LHC). The experiment shares the IP and the experimental cavern with the CMS experiment. TOTEM has performed various measurements on the forward diffracting pp processes. Among them, it precisely determined the pp cross sections at various center-of-mass energies $\sqrt{s}$, from 7 Tev up to 13 TeV \cite{TOTEM:AllCS}.  
The TOTEM experiment also performed a direct measurement of the $\rho$ parameter, the ratio between the real and the imaginary part of the forward scattering amplitude with 4-momentum $|t|=0$ \cite{TOTEM:rho8TeV}. The two sets of measurements, total cross section and $\rho$, lead to the first evidence of the Odderon \cite{TOTEM:OddFirst}, C-odd counterpart of the Pomeron, the latter being the main actor in the Regge theory for diffractive forward physics. Combing TOTEM and D0 data, collected at Fermilab in  $p\overline{p}$ collisions at $\sqrt{s}=1.96$ TeV, the discovery was claimed \cite{TOTEMD0}.\\
Differently from the other LHC experiments, TOTEM requires a special configuration of the machine optics to be able to detect elastic protons scattered at a few microradians. Such optics, characterized by a very high value of the beta function at the IP $\beta^*$ (from 90 m up to 6 km), requires dedicated LHC fills, referred to as \emph{special run}.
To better constrain the theoretical models on the Odderon, additional special runs were approved by the LHC, in particular one at $\sqrt{s}= 13.6$ TeV with $\beta^*$=90 m, aiming to determine the total cross section, scheduled for June 2023.\\
The precise measurement of the total cross section with the \emph{luminosity-independent} method \cite{TOTEM:7TeV} requires the simultaneous determination of the differential elastic cross section and the inelastic event rate.
The detection of very forward protons originating from elastic collisions is performed in the Roman Pots (RP), movable pipe insertions with a secondary vacuum, symmetrically located at around 200–220 m from the LHC IP5.
The RPs can be moved into the primary vacuum of the machine through vacuum bellows, allowing the equipped detector to approach the LHC beam down to few millimetres. In TOTEM, they are instrumented with silicon strip detectors \cite{TOTEM_std}.
Inelastic events are instead detected by dedicated detectors, called forward telescopes, symmetrically integrated with the CMS structure at few meters from the IP, instrumenting the forward pseudorapity $\eta$ region. For the previous measurements of TOTEM, two inelastic telescopes were in use, T1 and T2 \cite{TOTEM_std}. However, after having successfully sustained a radiation damage 10 times above the design requirements, they have been dismounted. An R\&D was hence initiate to replace the inelastic detector T2 with a new inelastic telescope, called nT2 \cite{nT2TDR}. The nT2 is designed to cover a pseudorapidity region 5.3 < $|\eta|$  < 6.5, allowing to detect more than 90\% of the inelastic events.

\section{The nT2 detector for inelastic event tagging}\label{sec:nT2}
\begin{figure}[htbp]
\centering % \begin{center}/\end{center} takes some additional vertical space
  \includegraphics[width=0.8\linewidth]{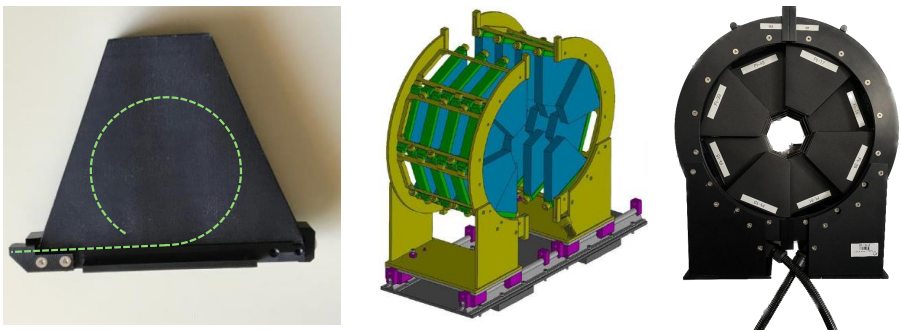}
\caption{\label{fig:detAss} On the left, a picture of one detector tile. The green dashed line represents the position of the WLS fiber. In the center, the design of one arm of the detector. On the right, a picture of one arm of the detector before the installation.}
\end{figure}

\begin{figure}[htbp]
\centering % \begin{center}/\end{center} takes some additional vertical space
  \includegraphics[width=0.80\linewidth]{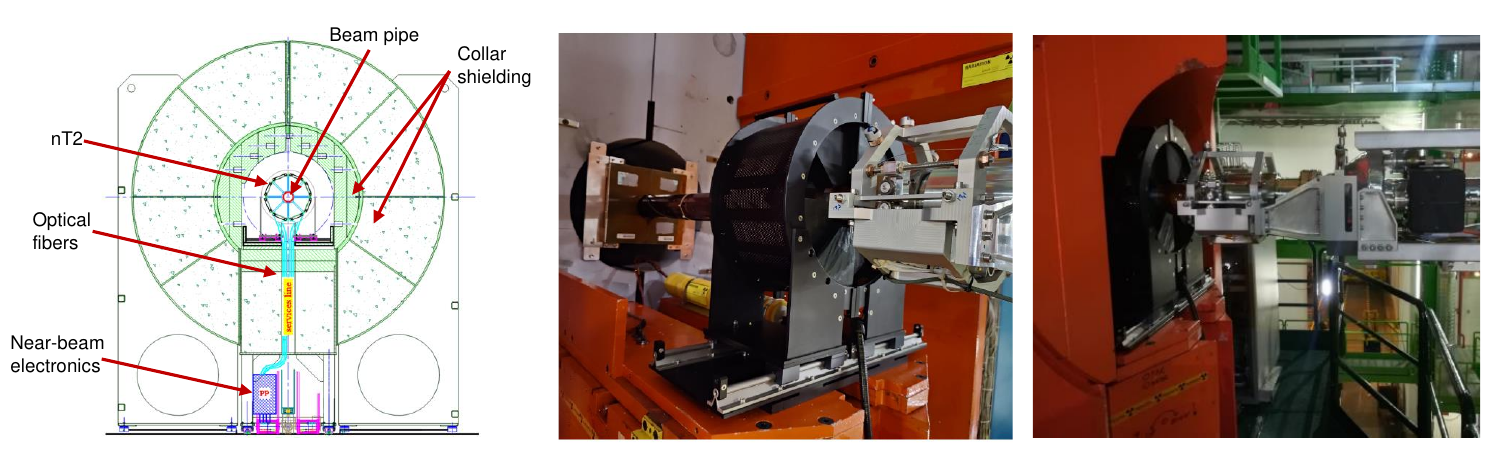}
\caption{\label{fig:detInst} On the left, a drawing of the detector integrated in CMS. The beam pipe is perpendicular to the picture, in the center of the nT2 detector. On the center and right, two pictures taken during the installation training with a mock-up.}
\end{figure}

To cope with the project budget, requirements and R\&D schedule, scintillators coupled to Silicon PhotoMultipliers (SiPM) were selected as the detection technology for the nT2.
Some tight constraints, derived from the operational environment and the LHC schedule, were taken into consideration during the design of the detector.
The installation was scheduled to happen during an LHC technical stop, a short period of a few days in between normal operations. The special run, with a foreseen duration of about 24 h, would have taken place only a couple of days after the installation, leaving a very short time for the detector commissioning and the eventual bug fixing. Moreover, the installation of the detector, symmetrically located at $\sim 15$ m from the IP close to the beam pipe, required the opening of the CMS forward radiation shields, exposing the technicians to a non negligible radiation dose. Hence, the detector was divided in two sections. The inner section, inside the radiation shields, had to be installed in minutes. After that, the shields must be closed and the inner part of the detector would no longer be accessible. The outer section, outside the shields, would still have been accessible, with some limitation, during the next two days, until the end of the technical stop.\\
The base unit of the nT2 is made of plastic scintillator tiles (EJ-204, manufactured by Eljen), with a thickness of 2 cm (figure \ref{fig:detAss}, left). The tiles are cut into a trapezoidal shape, with a total surface of $\sim100$ cm$^2$. On one face, a sigma-shaped groove is made to embed a WaveLength Shifter (WLS) fiber with a diameter of 1 mm. Each unit is then wrapped individually and secured to a mechanical support frame. A mechanical splicer is directly placed at the exit point of the WLS fiber.
The splicer connects, with a glue-less connection, the WLS fiber to a clear fiber of 2 m length to reduce the light attenuation from the sensor to the light detector. This design ensures that both the tiles and the external fibers can be independently and quickly replaced in case of failure. The nT2 is then assembled in quarters, two for each side of the interaction point (called \emph{arm}), designed to be closed around the beam pipe (figure \ref{fig:detAss}, center and right). The size of one arm of the detector, after being closed around the beam pipe, is $\sim 25$ cm in the direction along the beam axis, with an outer diameter of $\sim 50$ cm, including the external support structure. Each quarter is made up of 4 detection layers, each one formed by 4 tiles arranged on a semi-circle with a partial overlap. A set of 4 aligned tiles is called a \emph{wedge}. Hence, each particle will traverse at least 4 aligned tiles in the wedge, providing a good level of redundancy for particle detection and noise suppression. With such a design, the installation of the inner section of the detector on pre-installed rails can be done in few minutes and, after routing the fiber bundles through the available slot, the shields of CMS can be closed for a much safer environment to work in, leaving only passive components inside (figure \ref{fig:detInst}). 
The fibers are read out with an S13361-6050 16-channels SiPM matrix from Hamamatsu, one for each quarter, located outside the shields in the outer section of the detector together with the rest of the electronics. The MPPC bias is generated by temperature-compensated voltage converters, one for each matrix, controlled through a serial interface.
To digitize the signal the detector relies on a fast discriminator, the NINO chip \cite{NINO}, a differential amplifier and fixed threshold discriminator. An additional feature of the chip is that the digital output signal has a Time Duration (TD) that depends on the input charge Q, with higher charges generating longer TD.\\
For the read-out and control electronics, we benefit from the fruitful collaboration established between the TOTEM and CMS collaborations.
Indeed, in the last years the TOTEM read-out and control systems have been integrated in the CMS infrastructure.
This was made in the framework of the joint CMS-TOTEM Precision Proton Spectrometer project \cite{CTPPS_tdr} (CT-PPS, now only PPS).
The electronics from the timing system of the PPS detector \cite{TimingPPS} has been re-adapted to meet the requirements of the nT2 and branched to the previous T2 services. Hardware-wise, the only new component developed is represented by an interface mezzanine board, hosting the SiPM matrices, the HV controllers, the NINO chips and the digital-to-analog converters for the control of the discrimination thresholds. The mezzanine is then mounted on the CMS-TOTEM general purpose digitizer board, used by the timing system of PPS. The integration of the nT2 with the CMS control and read-out systems relies on the firmware loaded onto the SmartFusion2 System-on-Chip (SoC) FPGA on the digitizer board. In addition to event building, data buffering and communication with the nT2 mezzanine components, it also embeds a Time-to-Digital Converter (TDC) feature, adding a coarse timestamp (6.25 ns bin) to the leading and trailing edges of the NINO signals, used in the offline analysis. Several capabilities in terms of debugging and online monitoring have been implemented, in particular real-time efficiency and particle rate measurements.
Control of mezzanine devices, trough different serial protocols, is handled by the ARM CORTEX M3 integrated in the SoC FPGA, adding flexibility and fast debugging in case of need, even during the data taking. The C code for the microcontroller is stored in the integrated flash memory and it is rebootable through the firmware.

\section{Operations and preliminary performance}\label{sec:operation}

\begin{figure}[htbp]
\centering % \begin{center}/\end{center} takes some additional vertical space
  \includegraphics[width=0.30\linewidth]{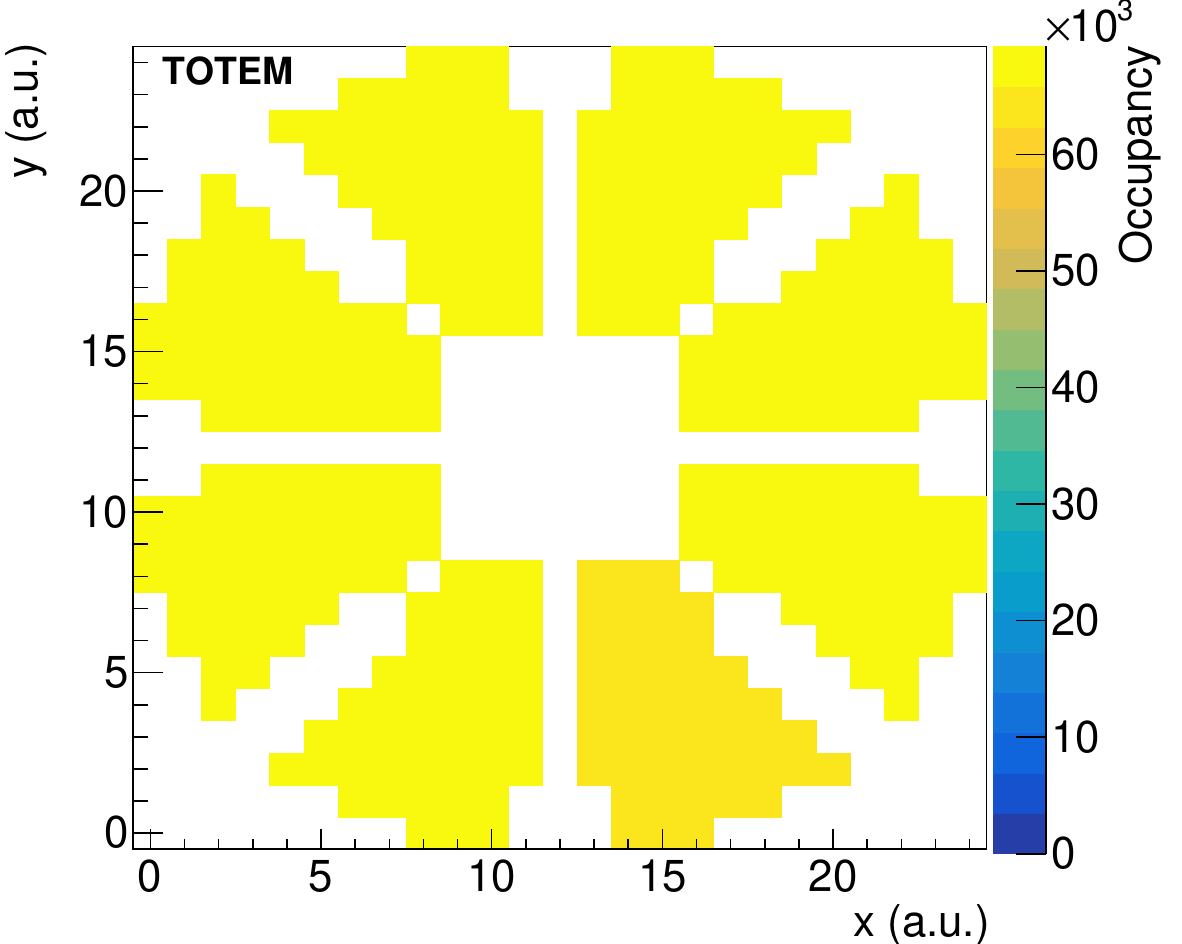}
  \includegraphics[width=0.25\linewidth]{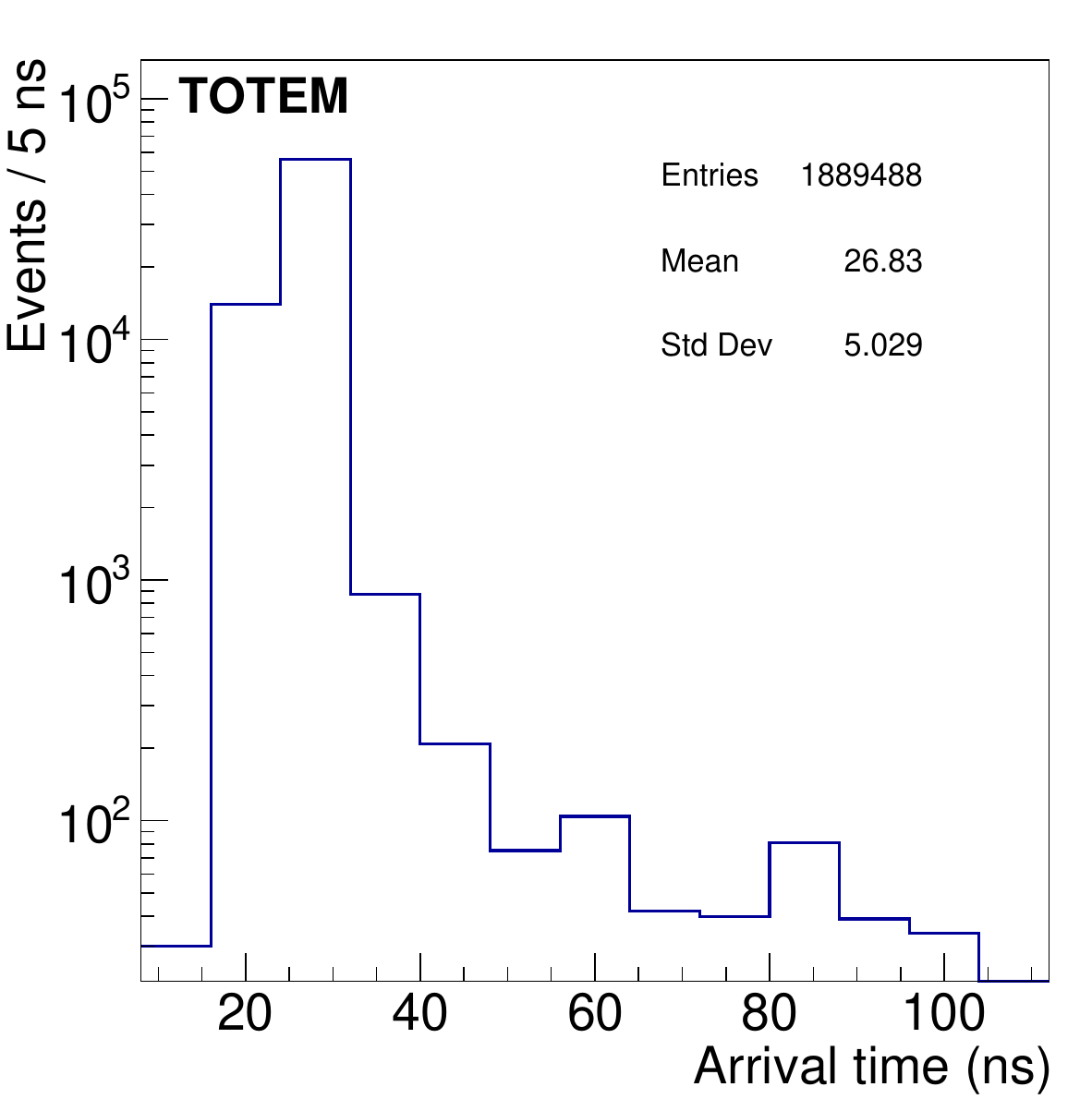}
  \includegraphics[width=0.25\linewidth]{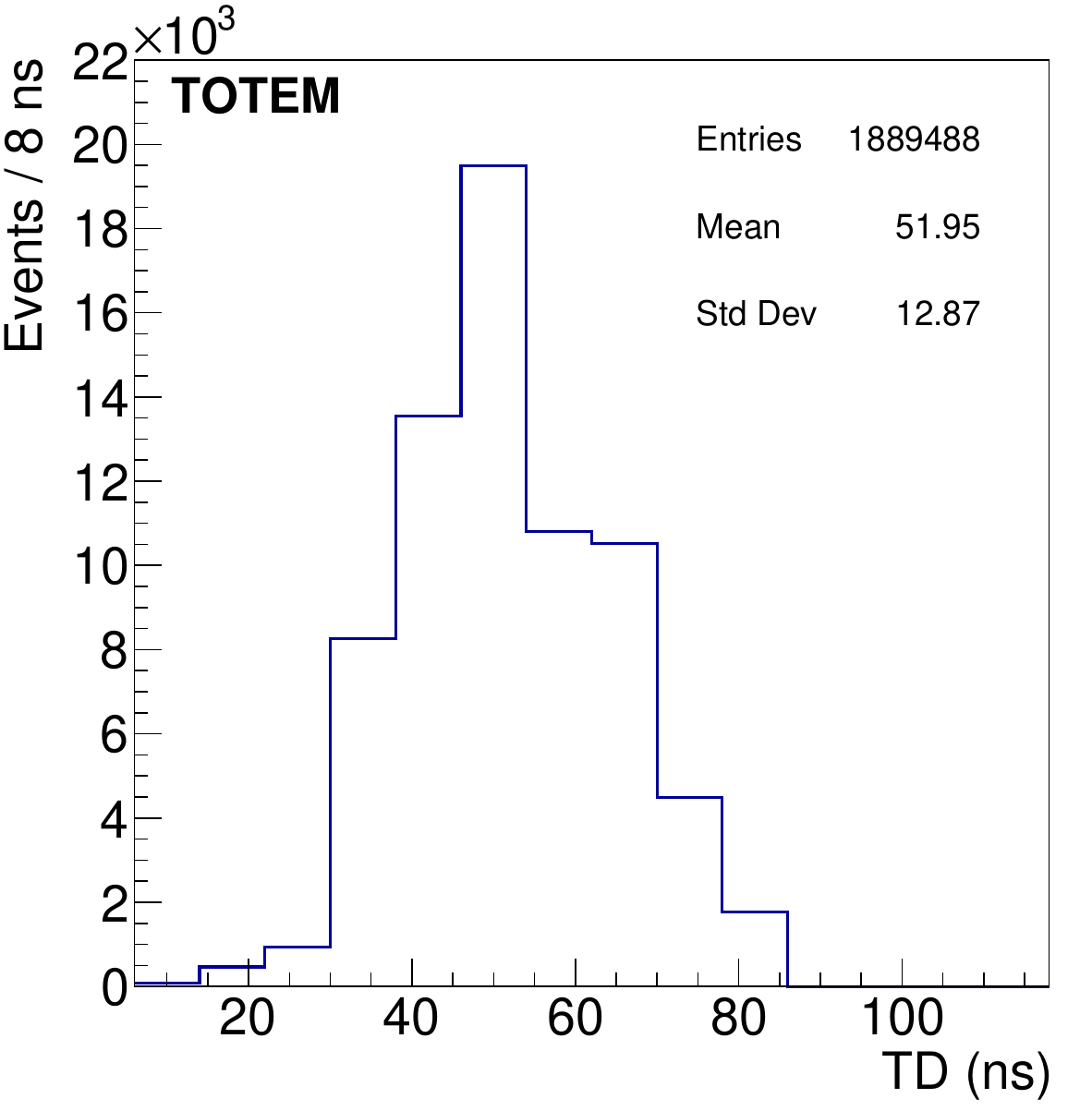}
\caption{\label{fig:occupancy} On the left, occupancy of the wedges of one arm of nT2. The bins are shaped to provide a simplified view of the real detector as seen from the IP. The histograms on the right side are examples of the signal arrival time and the TD for one channel of the detector.}
\end{figure}

\begin{figure}[htbp]
\centering % \begin{center}/\end{center} takes some additional vertical space
  \includegraphics[width=0.43\linewidth]{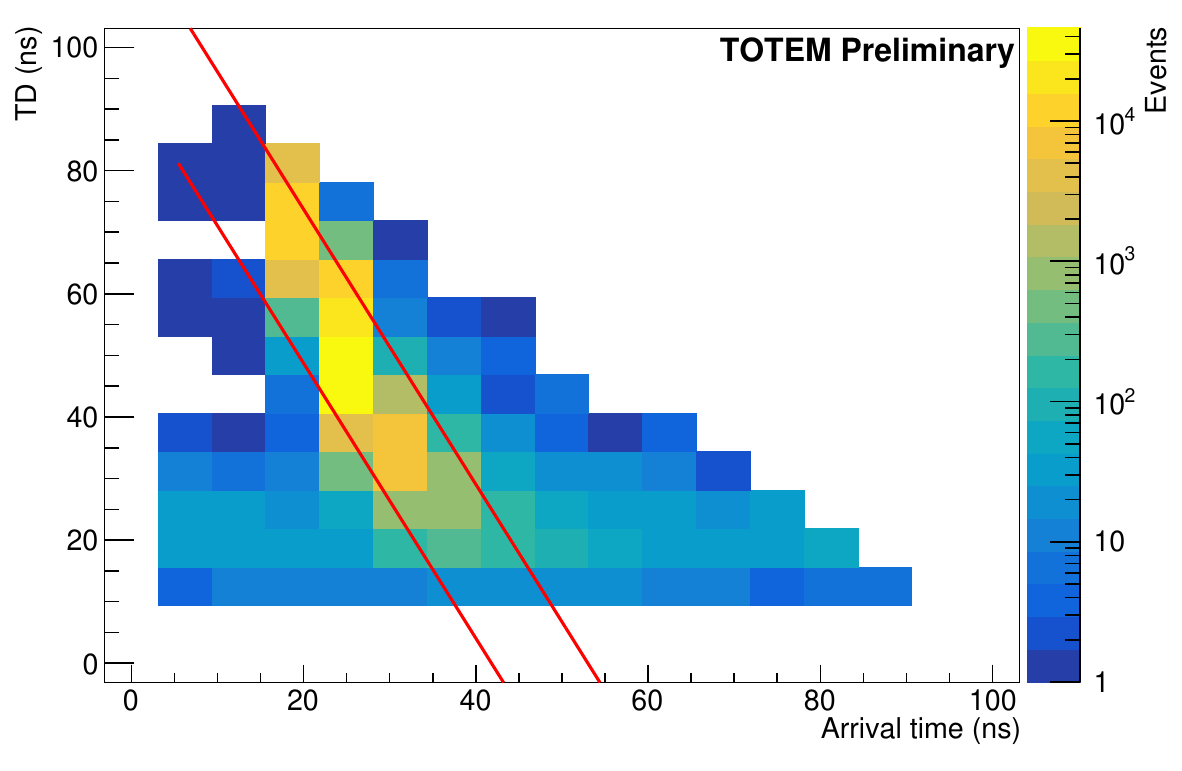}
  \includegraphics[width=0.41\linewidth]{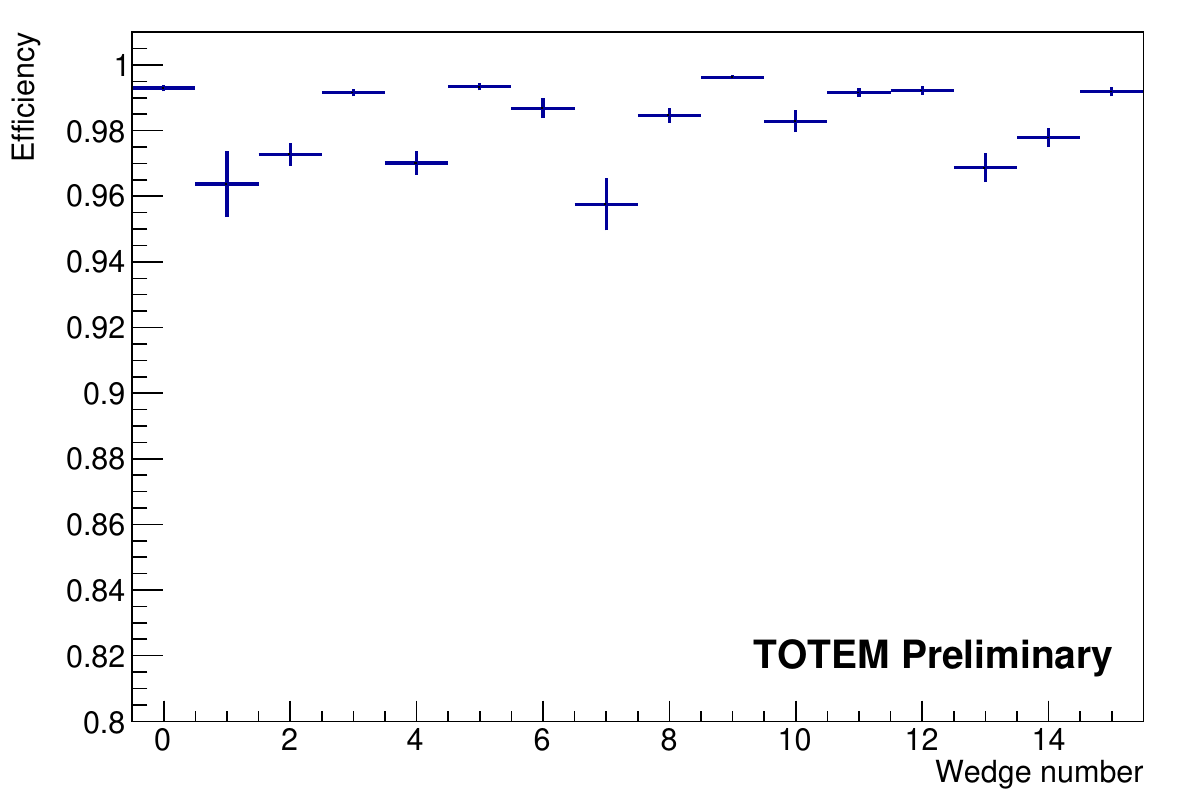}
\caption{\label{fig:preRes} On the left, correlation plot between the particle arrival time and the signal TD, with an example cut superimposed. On the right, the preliminary efficiency of all wedges of the detector.}
\end{figure}

The detector installation was performed within the foreseen timeline. On both sides of the IP the installation of the inner part of the detector was made in less than 10 minutes, after that the CMS forward shields were closed. The dose collected by the technician was negligible ($\sim10$ uSv).\\
The special run took place in June, few days after the detector installation and lasted for $\sim$ 30 hours.
Since the beginning of the data taking the detector was perfectly integrated with the CMS control and acquisition system, with all online reconstruction and monitoring software working as intended.
In figure \ref{fig:occupancy}, the occupancy in each wedge on one side of the IP, as detected by the online data quality monitoring, is shown. In there, a wedge is defined as active if there is a coincidence of at least three tiles. 
In the center and right side of figure \ref{fig:occupancy} are instead reported examples of the online histograms of the particle arrival times (defined as the timestamp of the signal leading edges) and the signal TDs respectively.\\
Data analysis is currently ongoing, with some preliminary results on the detector performance already available. On the left of figure \ref{fig:preRes}, an example of the correlation between the particle arrival time and the TD is reported for one of the channels. As expected, there is a negative correlation, with signals with higher TD having lower arrival times. This is expected as a signal with larger charge (and hence TD) goes above the fixed threshold of the discriminator before a simultaneous signal with a lower charge. This correlation is currently exploited to suppress noise hits by identifying an acceptance region in the correlation plot. In the figure an example cut is shown, defining a hit as good if it lies between the two red lines.
On the right of figure \ref{fig:preRes} we reported the preliminary results on the detector efficiency for each wedge. First, the efficiency of each tile has been evaluated by exploiting the large track multiplicity generated by inelastic events. By checking the presence of hits in the nearby wedges and in the other tiles of the wedge, it was possible to compute the efficiency of a given tile. The overall efficiency of a wedge is then determined with a combinatorial calculation, assuming all tiles have independent efficiency and defining a wedge as efficient if at least 2 tiles out of 4 have hits. All wedges show an efficiency of at least 96\%, with a median of 99.7\%. No noise suppression based on the hit time stamps is performed. If we apply a tight selection cut as shown in the left plot on all tiles, the median of the efficiency becomes 98.6\%. Work is ongoing to define the optimal cut on the noise. Wedges with lower efficiency can eventually be recovered by exploiting the azimuthal symmetry of inelastic events in the LHC collisions. 

\section{Conclusions}\label{sec:concl}
The measurements performed by the TOTEM experiment brought a significant improvement in the forward diffractive physics, culminating in the discovery of the Odderon. The LHC approved a special run to measure the total cross section at $\sqrt{s}=13.6$ TeV.
The precise measurement of the total cross section through the luminosity independent method needed a new inelastic telescope to be designed, based on scintillators read out by SiPMs.
The detector, to address the extremely tight constraint in terms of installation and commissioning time, has been divided in a passive inner part included in the CMS forward radiation shield and an outer part. Large part of the electronics has been readapted from the timing system of the PPS sub-detector of CMS, with some key components developed on purpose. The detector, with the inner part installed in less than 10 minutes, has been commissioned in the short time available and successfully operated for the special run of TOTEM in June 2023. Preliminary results demonstrate the offline capability to suppress residual noise hits based on the time measurements included in the data. The detector shows a high efficiency of all wedges, with an average value in the range 99.7\%-98.6\%, depending on the quality cut applied to the hits. 
Analysis of the detector performance is currently being finalized and the collected data will be then used to complete the TOTEM physics program.

\section*{Acknoledgment}
We would like to thank the CMS collaboration for their support during the detector installation and the data taking as well as for the analysis.
%\bibliographystyle{JHEP}
%\bibliography{Biblio}

\end{document}